\documentclass{ws-mpla}

\usepackage{epsf}
\usepackage{graphicx}
\usepackage{dcolumn}
\usepackage{bm}

\begin{document}

%
\catchline{}{}{}{}{}
%

\title{Parametrizations of the Dark Energy Density
and Scalar Potentials}

\author{\footnotesize Zong-Kuan Guo and Nobuyoshi Ohta}
\address{Department of Physics, Kinki University,
 Higashi-Osaka, Osaka 577-8502, Japan}
\author{Yuan-Zhong Zhang}
\address{{Institute of Theoretical Physics, Chinese Academy of
 Sciences, P.O. Box 2735, Beijing 100080, China.}\\
{guozk@phys.kindai.ac.jp}, {ohtan@phys.kindai.ac.jp}}

\maketitle


\begin{abstract}
We develop a theoretical method of constructing the scalar
(quintessence or phantom) potential directly from
the dimensionless dark energy function $X(z)$, the dark energy
density in units of its present value. We apply our method
to two parametrizations of the dark energy density,
the quiessence-Lambda ansatz and the generalized Chaplygin
gas model, and discuss some features of the constructed potentials.
\end{abstract}

\ccode{PACS Nos.: 98.80.Cq, 98.65.Dx}


Recent observations of type Ia supernovae suggest that the
expansion of the universe is accelerating and that two-thirds
of the total energy density exists in a dark energy component
with negative pressure.\cite{AGRSP1,AGRSP2}
In addition, measurements of the cosmic microwave
background\cite{DNS} and the galaxy power spectrum\cite{MT}
also indicate the existence of the dark energy.
The simplest candidate for the dark energy is a cosmological
constant $\Lambda$, which has pressure
$P_\Lambda=-\rho_\Lambda$. Specifically, a reliable model
should explain why the present amount of the dark
energy is so small compared with the fundamental scale
(fine-tuning problem) and why it is comparable with the
critical density today (coincidence problem).
The cosmological constant suffers from both these problems.
One possible approach to constructing a viable model for
dark energy is to associate it with a slowly evolving and
spatially homogeneous scalar field $\phi$, called
``quintessence''.\cite{RP1,RP2,CDS,PJS1,PJS2} Such a model
for a broad class of potentials can give the energy density
converging to its present value for a wide set of initial
conditions in the past and possess tracker behavior
(see, e.g., Ref.~\refcite{SS1,SS2,SS3,SS4} for reviews with more complete lists
of references).

The dark energy is characterized by its equation of state $w$ and
its energy density $\rho$, which are in general functions of
redshift $z$ in scalar field models. For a minimally coupled
scalar field model, the quintessence potential $V(\phi)$ may be
reconstructed from supernova observations\cite{SS1,DHMT1,DHMT2,DHMT3,DHMT5,DHMT6}
and directly from the effective equation of state function $w(z)$.\cite{GUO}
Given an evolution of the universe, the potential for a scalar and
tachyon field can be constructed.\cite{vit04} Moreover, the
cosmological evolutions may be drastically different for arbitrary
initial conditions on the tachyon field.\cite{vit04} Recently, the
reconstruction method has been studied in the k-essence
models,\cite{tsu05,LGZ1,LGZ2} tachyon models,\cite{hao02,tsu05} 5D
cosmological models\cite{XLP1,XLP2,XLP3} and scalar-tensor
theories.\cite{CNO1,CNO2}

Although the equation of state $w(z)$ has been generally chosen to
parametrize the dark energy, there are several reasons why the parametrization
by the energy density $\rho(z)$, which depends on its equation of state
$w(z)$ through an integral, gives a better one.
Firstly, it can be constrained more tightly than $w(z)$ for given
observational data.\cite{YWPMG1,YWPMG2} Secondly, parametrization of the energy
density provides a more flexible approach, which can determine the
properties of dark energy in a model independent manner.
Finally it was shown that assuming that
$w$ is constant or greater than $-1$ can lead to gross errors in
estimating the true equation of state.\cite{mao01}
This smearing effect can be decreased by using
the dark energy density $\rho(z)$. In this letter we develop a
theoretical method of constructing the scalar potential $V(\phi)$
directly from the dark energy density $\rho(z)$, which connects
the energy density parametrization to the physically effective
field theory. We apply this method to two parametrizations and
discuss some features of the resulting potentials.

Our method is new in that it relates directly the scalar
potential to the dark energy function and it is also extended
to the phantom field models. This enables us to
construct easily the potential without assuming its form.
For instance, in the reconstruction method proposed in
Refs.~\refcite{SS1,DHMT1,DHMT2,DHMT3,DHMT5,DHMT6}, the reconstruction
equations relate the potential and the equation of state to measurements
of the luminosity distance. The potential may thus be
reconstructed by way of the luminosity distance from
supernova data. Usually this can be done by assuming the
form of a potential.
Compared to the methods discussed in Refs.~\refcite{vit04} and \refcite{GUO},
it is extended to the construction of a phantom field potential.
A similar reconstruction of the potential was also considered
in Ref.~\refcite{CNO1,CNO2} using a single non-canomical scalar field
and the Hubble parameter parametrized by time. We generalize
this approach by including the contribution of the cold dark matter
as well. Moreover our method itself has the advantage of using
parametrization of Hubble parameter by redshift which is directly
observed, but that in terms of time is not an observed quantity.


We consider a spatially flat FRW universe which is dominated
by the non-relativistic matter and a spatially homogeneous
scalar field $\phi$. The Einstein equation can be written as
\begin{eqnarray}
H^2 &=& \frac{1}{3M_{pl}^2}(\rho_m+\rho_\phi), \\
\frac{\ddot{a}}{a} &=& -\frac{1}{6M_{pl}^2}(\rho_m + \rho_\phi + 3P_\phi),
\end{eqnarray}
where $M_{pl} \equiv (8\pi G)^{-1/2}$ is the reduced Planck mass,
$a$ is the scale factor, $H=\dot{a}/a$ is the Hubble parameter,
and $\rho_m$ is the matter density. The energy density
$\rho_\phi$ and pressure $P_\phi$ of the evolving scalar field
$\phi$ are given by
\begin{eqnarray}
\label{rhop}
\rho_{\phi} &=& \pm \frac{1}{2}\dot{\phi}^2+V(\phi), \\
P_{\phi}    &=& \pm \frac{1}{2}\dot{\phi}^2-V(\phi),
\label{pp}
\end{eqnarray}
respectively, where $V(\phi)$ is the scalar field potential.
The upper (lower) sign corresponds to a quintessence (phantom) field
in Eqs. (\ref{rhop}) and (\ref{pp}) and in what follows.
Using the Einstein equations and the expressions for $\rho_\phi$
and $P_\phi$, one can obtain
\begin{eqnarray}
V &=& 3M_{pl}^2 H^2 + M_{pl}^2 \dot{H} - \frac{1}{2} \rho_m, \\
\dot{\phi}^2 &=& \mp \left(2M_{pl}^2 \dot{H} + \rho_m \right),
\end{eqnarray}
which can be rewritten as
\begin{eqnarray}
\label{vz}
V(z) &=& 3M_{pl}^2 H^2 - M_{pl}^2 (1+z) H \frac{dH}{dz}
 - \frac{1}{2} \rho_{m0}(1+z)^3, \\
\pm \left(\frac{d\phi}{dz}\right)^2 &=&
2M_{pl}^2 (1+z)^{-1}H^{-1}\frac{dH}{dz}
 -\rho_{m0}(1+z)H^{-2},
\label{dpz}
\end{eqnarray}
in terms of the redshift $z=-1+a_0/a$, and we have used the relation
$\rho_m =\rho_{m0}(1+z)^3$. Throughout this paper
quantities with subscript $0$ denote the values at the
redshift $z=0$ (present).

We define dimensionless quantities
\begin{equation}
\tilde{V} \equiv V / \rho_0, \quad
\tilde{\phi} \equiv \phi / M_{pl},
\end{equation}
where $\rho_0=\rho_{\phi0}+\rho_{m0}$ is the total energy
density at present time.
The construction equations (\ref{vz}) and (\ref{dpz}) can
then be written as
\begin{eqnarray}
\label{tvz}
\tilde{V}(z) &=& (1-\Omega_{m0})
\left[X(z)-\frac{1}{6}(1+z)\frac{dX(z)}{dz}\right], \\
\left(\frac{d\tilde{\phi}}{dz}\right)^2 &=&
\pm (1-\Omega_{m0})\frac{1}{(1+z)E^2(z)}\frac{dX(z)}{dz},
\label{tpz}
\end{eqnarray}
where $X(z) \equiv \rho_\phi(z)/\rho_{\phi 0}$ is the
dimensionless dark energy function, the dark energy density
in units of its present value,
$\Omega_{m0} \equiv \rho_{m0}/\rho_0$ is the present day
density parameter of matter, and
\begin{equation}
\label{ez}
E(z) \equiv
\left[\Omega_{m0}(1+z)^3+(1-\Omega_{m0})X(z)\right]^{1/2}
\end{equation}
is the cosmic expansion rate relative to its present value.
Given the dark energy function $X(z)$,
the construction equations (\ref{tvz})
and (\ref{tpz}) will allow us to construct the scalar
potential $V(\phi)$, quintessence potential for $dX/dz>0$
and phantom potential for $dX/dz<0$.
The sign of $d\phi/dz$ in fact is
arbitrary, as it can be changed by the field redefinition,
$\phi \to -\phi$. So we choose $d\phi/dz < 0$ in the
following discussions, that is, the scalar field $\phi$
increases as the universe expands.


As examples, let us now consider the following two
parametrizations: the quiessence-$\Lambda$ ansatz and the
generalized Chaplygin gas model.
\vspace{2mm}

{\noindent \bf (i) Quiessence-$\Lambda$ ansatz:}
\begin{equation}
\label{case1}
X(z) = A + (1-A)(1+z)^{3(1+w)},
\end{equation}
where $A$ and $w$ are constant.
This ansatz involves the combination of cosmological constant
with ``quiessence'' (dark energy with constant equation of state
parameter), called quiessence-$\Lambda$ ansatz.\cite{WT1,WT2}
Both early on and in the distant future, the dark energy
approaches either a constant equation of state $w$ or a constant
density, depending on the sign of $(1+w)$.
When the universe is dominated by dark energy, we obtain
\begin{eqnarray}
\label{qlp}
\tilde{V}(\tilde{\phi}) = A(1-\Omega_{m0})
\left[1+\frac{1-w}{2}
 \sinh^2\frac{\sqrt{\pm 3 (1+w)}(\tilde{\phi}-\tilde{\phi}_0)}{2}
 \right].
\end{eqnarray}
For $w < -1$, this result is consistent with that
in Ref.~\refcite{vit04}.
Note that $d\tilde{V}/d\tilde{\phi} < 0$ for $w > -1$ and
$d\tilde{V}/d\tilde{\phi} > 0$ for $w < -1$.
This result implies that the quintessence field rolls down
its potential while the phantom field climbs up its potential.
\vspace{2mm}

{\noindent \bf (ii) Generalized Chaplygin gas model:}
\begin{equation}
X(z) = \left[A_s + (1-A_s)(1+z)^{3(1+\alpha)}\right]^{\frac{1}{1+\alpha}},
\end{equation}
where $A_s$ and $\alpha$ are constant. This ansatz can be obtained by
modeling the dark energy as generalized Chaplygin gas with an equation
of state $P_c=-A/\rho_c$.\cite{KMP,bil021,bil022}
By choosing different ranges for
the parameters ($A_s$,$\alpha$), this ansatz can behave as standard
dark energy model with asymptotic de-Sitter phase, early phantom model
in which $w \ll -1$ at early times and asymptotically approaches
$w = -1$ at late times, late phantom model with $w \approx -1$ at early
times and $w \to - \infty$ at late times, and transient model where
the present acceleration of the universe is only temporary as it again
enters the dust-dominated decelerating phase in future.\cite{AASS}
When the universe is dominated by the quintessence-like Chaplygin gas
with $A_s < 1$, one gets
\begin{eqnarray}
\label{chap}
\tilde{V}(\tilde{\phi}) &=&
 \frac{1}{2}A_s^{\frac{1}{1+\alpha}}(1-\Omega_{m0})
 \left[\cosh^{\frac{2}{1+\alpha}}
 \frac{\sqrt{3}(1+\alpha)(\tilde{\phi}-\tilde{\phi}_0)}{2}
 \right. \nonumber \\
 &&+ \left.\cosh^{\frac{-2\alpha}{1+\alpha}}
 \frac{\sqrt{3}(1+\alpha)(\tilde{\phi}-\tilde{\phi}_0)}{2}\right].
\end{eqnarray}
The potential (\ref{chap}) with $\alpha=1$ corresponds to
the original Chaplygin gas model, which was reconstructed
in Ref.~\refcite{KMP}. The generalized model with $\alpha=0$
give the same reconstructed potential as the
quiessence-$\Lambda$ model with $w=0$.
For the phantom-like Chaplygin gas with $A_s > 1$, we obtain
\begin{eqnarray}
\tilde{V}(\tilde{\phi}) &=&
 \frac{1}{2}A_s^{\frac{1}{1+\alpha}}(1-\Omega_{m0})
 \left[\sin^{\frac{2}{1+\alpha}}
 \frac{\sqrt{3}(1+\alpha)(\tilde{\phi}-\tilde{\phi}_0)}{2}
 \right. \nonumber \\
 &&+ \left.\sin^{\frac{-2\alpha}{1+\alpha}}
 \frac{\sqrt{3}(1+\alpha)(\tilde{\phi}-\tilde{\phi}_0)}{2}\right].
\end{eqnarray}

\begin{figure}
\begin{center}
\includegraphics[width=8cm]{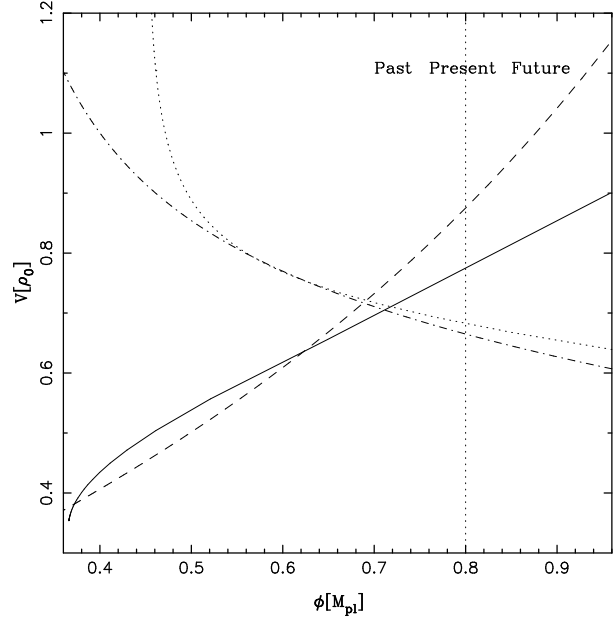}
\caption{Constructed scalar potentials for the quiessence-$\Lambda$
ansatz with $\Omega_{m0} = 0.3$ and $A = 0.5$. The solid, dashed, dotted
and dot-dashed lines correspond to $w = -1.4$, $w = -2.0$, $w = -0.9$ and
$w = -0.8$ respectively.}
\label{fig:quiessence}
\end{center}
\end{figure}

\begin{figure}
\begin{center}
\includegraphics[width=8cm]{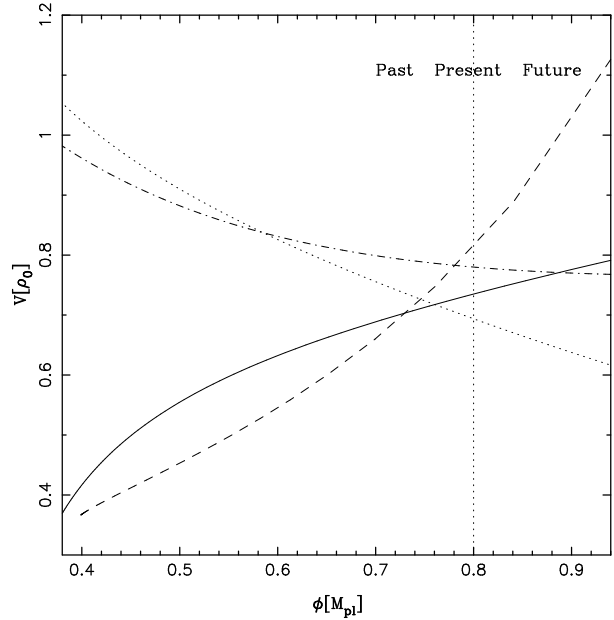}
\caption{Constructed scalar potentials for the generalized Chaplygin
gas model. The solid, dashed, dotted and dot-dashed lines represent
the early phantom model ($\Omega_{m0} = 0.3$, $A_s = 1.1$, $\alpha = -0.8$),
late phantom model ($\Omega_{m0} =0.35$, $A_s=1.5$, $\alpha=-1.7$),
transient model ($\Omega_{m0}=0.25$, $A_s=0.85$, $\alpha=-1.2$) and
standard model ($\Omega_{m0}=0.2$, $A_s=0.95$, $\alpha=0.2$)
respectively.}
\label{fig:chaplygin}
\end{center}
\end{figure}

In Fig.~\ref{fig:quiessence}, we have plotted the constructed
quintessence or phantom potential for the quiessence-$\Lambda$
ansatz with $\Omega_{m0}=0.3$ and $A=0.5$. Clearly, Eq.~(\ref{case1})
indicates that the effective equation of state approaches $-1$
if $w>-1$ and $w$ if $w<-1$.
As shown in Fig.~\ref{fig:quiessence}, in the case of $w>-1$ the
constructed quintessence potentials are in the form of a runaway
type, in agreement with our previous analysis.\cite{GUO}
On the other hand, for $w<-1$ the potentials increase as the universe
expands.
The asymptotic form is obtained as follows.
The phantom energy density increases with time,
so the matter component in Eq.~(\ref{ez}) and the
first term in Eq.~(\ref{case1}) become negligible when $z \to -1$.
We find that the potentials tend to the exponential form:
\begin{eqnarray}
\tilde{V}(\tilde{\phi})
= \frac{1}{2}(1-\Omega_{m0})(1-A)(1-w)
 \exp\left[\sqrt{-3(1+w)}(\tilde{\phi}-\tilde{\phi}_0)\right],
\end{eqnarray}
which leads to unwanted future singularity called ``Big Rip".
Therefore, in this parametrization~(\ref{case1}), the universes
either accelerates forever (if $w>-1$) or reaches a Big Rip (if
$w<-1$) at late times. In the former case, the
quiessence-dominated phase is transient and de-Sitter is a
late-time attractor. In the latter case, the phantom dominates the
universe in the late times. Thus a potential with a local maximum
can not be constructed, in which case the phantom dark energy is
transient and de-Sitter is a late-time attractor.~\cite{car031,car032}

Fig.~\ref{fig:chaplygin} shows the constructed scalar potentials for
the generalized Chaplygin gas model with $\Omega_{m0}=0.3$.
One can see that the potentials tend to be flat for the standard and
early phantom models while the potentials become steep
as the universe expands for the transient and late phantom models.

In the evaluation of these equations, we have also chosen the
initial values of the scalar field $\tilde \phi_0=0.8$
at the redshift $z=0$ (present).
The value of $\tilde \phi_0$ is chosen for the purpose of
definiteness. If we shift its value, it simply results in the
shift of the value of the scalar field; the potential in
Figs.~\ref{fig:quiessence} and \ref{fig:chaplygin}
is shifted horizontally. It has no influence on the evolution
of the universe and the shape of the quintessence potential.


Compared to the method proposed in Ref.~\refcite{GUO}, it is
extended to the phantom field with a negative kinetic term.
The dark energy density function, $X(z)$, is related to $w(z)$
as follows:
\begin{equation}
X(z)=\exp\left[3\int_{0}^{z}(1+w)d\ln(1+z)\right],
\end{equation}
so that $w=-1+\frac{1}{3}(1+z)\,d\ln X/dz$.
One can see that it is easier to extract $X(z)$ from the data
than to extract $w(z)$. Wang and Garnavich argued that $X(z)$
should be preferred since it suffers less from the smearing effect
that makes constraining $w(z)$ extremely difficult.\cite{YWPMG1,YWPMG2}
By precision measure of $X(z)$, we can obtain the scalar potential
using the construction method.

In conclusion, we have developed a method of constructing
the quintessence and phantom potentials directly from the dark
energy density function $X(z)$. From the theoretical viewpoint,
the method connects the energy density parametrization
to the effective field theory and throws light on the physical
nature of dark energy. Then we have considered two
parametrizations and constructed the scalar potential.

It is emphasized that the method is invalid when the
effective equation of state crosses the phantom divide
$w=-1$ since neither a single quintessence field
nor a single phantom field realizes the crossing of
the phantom divide.\cite{RRC} But in the quintom model with a quintessence
field and a phantom field this case can be realized easily.\cite{ZXM1,ZXM2}

\section*{Acknowledgments}

The work of NO was supported in part by the Grant-in-Aid for
Scientific Research Fund of the JSPS No. 16540250.
YZZ was in part supported by National Basic Research
Program of China under Grant No. 2003CB716300 and
by NNSFC under Grant No. 90403032.



\end{document}